\documentclass[12pt,onecolumn,oneside,a4paper,peerreview]{IEEEtran_kit}

\usepackage{cite}
\usepackage{float}
\usepackage{graphicx}
\usepackage{amsmath}
\usepackage{times}
\usepackage{graphicx}
\usepackage{latexsym}
\usepackage{graphicx}
\usepackage{bm}
\usepackage{amssymb}
\usepackage[center]{caption2}
\usepackage{stfloats}
\usepackage{cases}
\usepackage{array}
\usepackage{setspace}
\usepackage{fancyhdr}
\usepackage{color}
\usepackage{subfigure}
\usepackage{citesort}
\usepackage{multirow}
\usepackage{amsmath}
\usepackage{cases}
\usepackage[noend]{algpseudocode}
\usepackage{algorithm}
\usepackage{algorithmicx}

\usepackage{diagbox}
\usepackage{algpseudocode}
\usepackage{epstopdf}
\usepackage{varwidth}
\usepackage{pifont}

\def\bm#1{\mbox{\boldmath $#1$}}

\renewcommand{\baselinestretch}{0.97}

\begin{document}
\title{Deep Learning based Channel Estimation for Massive MIMO with Hybrid Transceivers}
 \author{\authorblockN{Jiabao Gao, Caijun Zhong, Geoffrey Ye Li, and Zhaoyang Zhang
 \thanks{J. Gao, C. Zhong, and Z. Zhang are with the College of Information Science and Electronic Engineering, Zhejiang University, Hangzhou, China (Email:  \{gao\_jiabao, caijunzhong, ning\_ming\}@zju.edu.cn). Geoffrey Ye Li is with the Faculty of Engineering, Department of Electrical and Electronic Engineering, Imperial College London, England (Email: Geoffrey.Li@imperial.ac.uk).}
 }}
\maketitle
\begin{abstract}
Accurate and efficient estimation of the high dimensional channels is one of the critical challenges for practical applications of massive multiple-input multiple-output (MIMO). In the context of hybrid analog-digital (HAD) transceivers, channel estimation becomes even more complicated due to information loss caused by limited radio-frequency chains. The conventional compressive sensing (CS) algorithms usually suffer from unsatisfactory performance and high computational complexity. In this paper, we propose a novel deep learning (DL) based framework for uplink channel estimation in HAD massive MIMO systems. To better exploit the sparsity structure of channels in the angular domain, a novel angular space segmentation method is proposed, where the entire angular space is segmented into many small regions and a dedicated neural network is trained offline for each region. During online testing, the most suitable network is selected based on the information from the global positioning system. Inside each neural network, the region-specific measurement matrix and channel estimator are jointly optimized, which not only improves the signal measurement efficiency, but also enhances the channel estimation capability. Simulation results show that the proposed approach significantly outperforms the state-of-the-art CS algorithms in terms of estimation performance and computational complexity. 

\vspace{1cm}
\begin{center}
{\bf Index Terms}
\end{center}
Massive MIMO, channel estimation, hybrid analog-digital, angular space segmentation, deep learning.
\end{abstract}

\newpage

\section{Introduction}
Massive multiple-input multiple-output (MIMO) has many advantages in terms of spectral and energy efficiency and has been envisioned as one of the key enabling technologies in the fifth generation of wireless communication systems\cite{massiveMIMO1,massiveMIMO2}. However, the realization of the potential gains of massive MIMO systems heavily relies on the availability of accurate channel state information (CSI). In time-division duplex (TDD) systems, only the uplink channel needs to be estimated thanks to the channel reciprocity\cite{Reciprocity} and the training overhead scales linearly with the number of users, which is usually acceptable. However, in frequency-division duplex (FDD) systems, the downlink CSI needs to be estimated and fed back to the base station (BS) by the users, where the downlink training and uplink feedback overhead scales linearly with the number of antennas at the BS and can substantially deteriorate the system efficiency. 

Since the challenge mainly comes from the large number of antennas, dimension reduction is a natural idea that comes to mind. In practice, the BS is usually located in high altitude with few surrounding scatters\cite{narrow_AS}, so the angular spread of incident signals of each user at the BS is narrow. Consequently, the channel covariance matrix (CCM) possesses low-rank characteristics. To exploit this, the original channels are approximated with a few main eigenvalues and eigenvectors in \cite{LowRank1,LowRank2} to reduce the effective channel dimensionality. However, the involved eigen-decomposition operation requires high computational complexity and the acquisition of accurate CCM in massive MIMO systems also requires extra overhead. An alternative method is to exploit the basis expansion model (BEM), which reduces the number of parameters to be estimated by exploiting the channel sparsity in specific domains\cite{BEM3,BEM4}. In \cite{BEM3}, a spatial BEM has been proposed to transform the problem of estimating channel impulse responses to that of estimating spatial basis function weights, which are sparse due to the physical scattering characteristics. The spatial and frequency wideband effects are considered in \cite{BEM4}, where the channel sparsity in the angle and the delay domains is exploited, and angular and delay rotations are used to further enhance the sparsity level. Although more computationally efficient, the BEM methods inevitably introduce approximation error to channel estimation due to the imperfect model. A comprehensive overview of low-rank channel estimation methods for massive MIMO systems can be found in \cite{overview}.

In conventional massive MIMO systems, each antenna is equipped with a dedicated radio-frequency (RF) chain, which leads to high hardware and energy cost when the number of antennas is large. To tackle this issue, the so-called hybrid analog-digital (HAD) architecture has been proposed, where the multi-antenna array is connected to only a limited number of RF chains through phase shifters in the analog domain\cite{HAD1,HAD2}. However, the channel estimation problem becomes more difficult in the context of HAD since now the received signals at the BS are not the original signals at antennas, but only a few of their linear combinations. In this situation, the conventional least-square (LS) estimator becomes inefficient with dramatically increased overhead\cite{LS}. In \cite{HAD_angular}, the complete channels are obtained by LS in the preamble stage and directions-of-arrival (DoAs) of channel paths are estimated first. Since the DoAs change slowly and can be used for a relatively long period, only channel gains of each path need to be re-estimated. Usually, the number of paths is much smaller than that of antennas in millimeter wave systems, therefore greatly reducing the estimation overhead. An alternative method is to adopt the compressive sensing (CS) methods to directly recover the sparse channels all at once, such as orthogonal matching pursuit (OMP)\cite{HAD_CS1}, sparse Bayesian learning (SBL)\cite{HAD_CS2}, etc. Through embedding the structural characteristics of channel sparsity, several improved CS algorithms have been further proposed, including structured SBL\cite{Structured_SBL} and structured variational Bayesian inference (S-VBI)\cite{VBI}. However, the performance of the CS algorithms heavily relies on the channel sparsity and the computational complexity is relatively high.

The aforementioned methods either suffer from unsatisfactory performance or high complexity, hence channel estimation algorithms with better performance-complexity tradeoffs are urgently required for practical HAD massive MIMO systems. Recently, deep learning (DL) has been successfully applied to many areas in wireless communication\cite{survey,new_survey1,new_survey2}, including spectrum sensing\cite{SpectrumSensing}, resource management\cite{ResourceManagement1,ResourceManagement2,ResourceManagement3}, beamforming\cite{Beamforming1,Beamforming2,Beamforming3}, signal detection\cite{Powerof,model_driven,NewRef_SD}, and channel estimation\cite{DL_CE1,LDAMP,DL_CE2,DL_CE3,DL_CE4,ATT_CE1,ATT_CE2}. Thanks to the simple forward computation and the acceleration brought by dedicated hardware like graphics processing unit (GPU), the DL-based approaches are usually more computationally efficient than the conventional iterative optimization algorithms\cite{ResourceManagement2}. For DL-based channel estimation in \cite{DL_CE1}, a deep convolutional neural network (CNN) has been proposed to refine the coarse HAD massive MIMO channel estimation and the correlations of channels in the frequency and the time domains are exploited to further improve the estimation performance and reduce the overhead. In \cite{LDAMP}, a denoising-based approximate message passing network, originated from image recovery, has been used to exploit the sparsity of millimeter wave channel and improve estimation performance. In \cite{DL_CE2} and \cite{DL_CE3}, the pilot and channel estimator are jointly optimized for downlink massive MIMO with autoencoders, where the pilot matrix and channel estimator are modeled as the encoder and decoder, respectively. In \cite{ATT_CE1} and \cite{ATT_CE2}, different attention modules are inserted to the deep neural networks to better exploit the channel features and improve estimation performance. Apart from channel estimation, DL can also be applied to channel tracking and prediction. In \cite{DL_CE4}, a channel tracking method based on graph neural network has been proposed, which can better extract the spatial correlation of massive MIMO channels than other network architectures. In \cite{DL_CE5}, future channels are predicted based on the past estimations with a recurrent neural network while DL is used to obtain channels conveniently in \cite{DL_CE6} based on related environmental factors, including frequency, location, temperature, etc. In \cite{DL_CE7}, a sparse complex network has been developed to learn the FDD uplink-to-downlink mapping and directly predict downlink channels based on uplink channels.

In this paper, we propose a novel DL-based channel estimation framework. With the aid of global positioning system (GPS) information, users at different angles relative to the BS are divided into different regions, and their channels can be estimated by the most matching sensing and estimation processes realized by different deep neural networks. The main contributions can be summarized as follows:

\begin{itemize}
\item A novel angular space segmentation method is proposed to effectively exploit the spatial-clustered sparsity structure of angular domain channels. Specifically, the entire angular space is segmented into many small angular regions. For each region, a dedicated neural network is trained offline. When deployed online, the most suitable network is selected based on the GPS information of the user.

\item A deep neural network is designed for channel estimation in HAD massive MIMO systems. Adopting the autoencoder architecture, the region-specific measurement matrix is jointly learned with the channel estimator. The learned measurement matrix significantly improves the signal measurement efficiency than the conventional one, and the learned channel estimator has stronger estimation capability than the state-of-the-art CS algorithms, as well as lower computational complexity.
\end{itemize}

The remaining of this paper is organized as follows. In Section II, the considered HAD massive MIMO system model and the formulation of the uplink channel estimation problem is introduced. Section III presents the proposed DL-based channel estimation framework, including the overview of working procedures, the introduction to the angular space segmentation method, and the detailed network architecture design. Simulation results are provided in Section IV to validate the superiority of the proposed approach. Eventually, the paper is concluded in Section V.

Here are some notations used in this paper. We use italic, bold-face lower-case and bold-face upper-case letter to denote scalar, vector, and matrix, respectively. ${\left\| {\bf{A}} \right\|}$, ${\bf A}^{T}$, and ${\bf A}^{H}$ denote the $l$-2 norm, transpose, and Hermitian transpose of matrix $\bm{A}$, respectively. $\bm{I}$ denotes the identity matrix. $[\bm{A}]_{i,j}$ denotes the element at the $i$-th row and $j$-th column of matrix $\bm{A}$. $|a|$ denotes the modulus of complex number $a$. ${\mathbb C^{x \times y}}$ and ${\mathbb R^{x \times y}}$ denote the ${x \times y}$ complex and real spaces, respectively. $\mathcal{CN}(\mu,\sigma^2)$ denotes the distribution of a circularly symmetric complex Gaussian random variable with mean $\mu$ and covariance $\sigma^2$. $\mathcal{U}[a,b]$ denotes the uniform distribution between $a$ and $b$. $\lceil \cdot \rceil$ denotes the operation of rounding up a number to the nearest integer.

\section{System model and problem formulation}
In this section, the considered system model and channel model are introduced first. Then, the HAD massive MIMO channel estimation problem is formulated.

\subsection{System Model}
In this paper, we consider a single-cell massive MIMO system as illustrated in Fig. 1, where the BS equipped with an $N$-antenna uniform linear array (ULA) serves a user equipped with an $M$-antenna ULA. As in \cite{HAD1,VBI}, to reduce the hardware cost and circuit energy consumption, the HAD architecture is adopted, where the BS only has $R$ ($R\ll N$) RF chains. We will first consider the single-user case and discuss the extension to the multi-user case later.
\begin{figure}[htbp]
\centering
\includegraphics[width=1\textwidth]{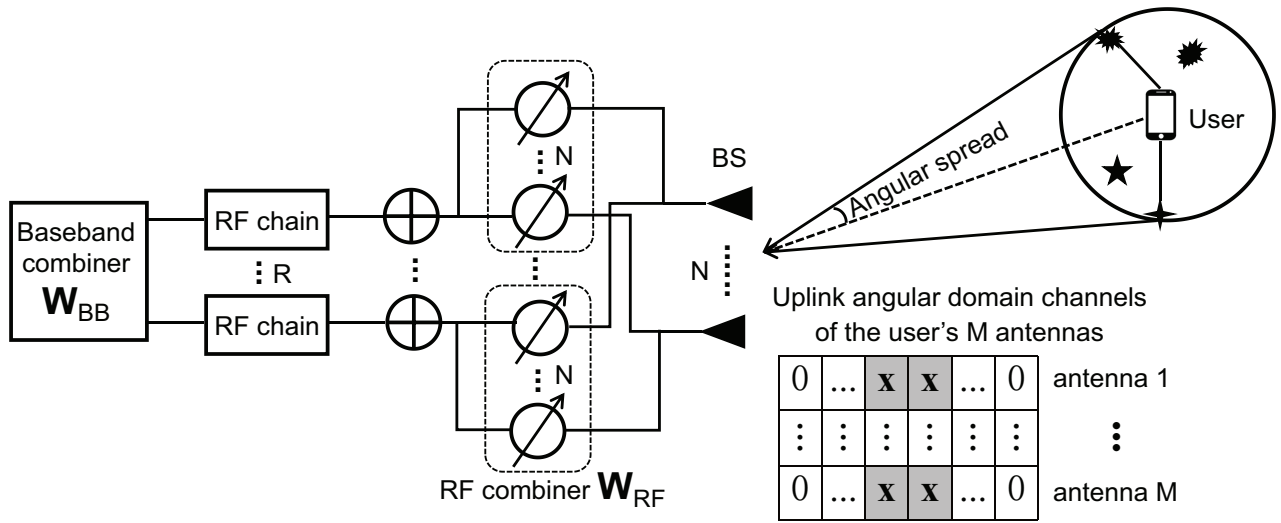}
\caption{Single-cell HAD massive MIMO system.}
\label{system}
\end{figure}

\subsection{Channel Model}
We consider TDD networks, where the downlink channel can be obtained by uplink channel according to channel reciprocity. Therefore, only the uplink channel needs to be estimated. Denote the uplink channel from the user to the BS as
\begin{equation}
\bm{H}=\frac{1}{\sqrt{N_p}}\sum_{i=1}^{N_p}\alpha_i\bm{a}_B(\theta_i)\bm{a}_U(\phi_i)^T\in \mathbb{C}^{N \times M},
\label{channel}
\end{equation}
where $N_p$ is the number of paths, $\alpha_i\sim \mathcal{CN}(0,1)$ denotes the complex gain of the $i$-th path, and $\theta_i$ and $\phi_i$ denote the angle-of-arrival (AoA) at the BS and angle-of-departure (AoD) at the user of the $i$-th path, respectively. For a ULA with half-wavelength among adjacent antennas, the steering vectors at the BS and the user can be expressed as $\bm{a}_B(\theta)=[1,e^{j\pi\sin(\theta)},\cdots,e^{j\pi\sin(\theta)(N-1)}]^T\in \mathbb{C}^{N \times 1}$ and $\bm{a}_U(\phi)=[1,e^{j\pi\sin(\phi)},\cdots,e^{j\pi\sin(\phi)(M-1)}]^T\in \mathbb{C}^{M \times 1}$, respectively.

Denote the azimuth angle of the user relative to the BS as $\theta_{az}$, and the angular spread of the user's channel paths as $\bigtriangleup_{\theta}$, we assume $\theta_i\sim \mathcal{U}[\theta_{az}-\bigtriangleup_{\theta},\theta_{az}+\bigtriangleup_{\theta}]$. As in \cite{BEM3,VBI}, the narrow angular spread assumption is adopted. Therefore, $\bigtriangleup_{\theta} \ll \pi$ as in practical environments. On the other hand, scatters are assumed to be randomly distributed around the user, that is, $\phi_i\sim \mathcal{U}[0,2\pi]$. To better exploit the channel features, we first convert the original channels into the angular domain by
\begin{align}
\bm{X}=\bm{B}^H\bm{H}\in \mathbb{C}^{N \times M},
\end{align}
where $\bm{X}$ denotes the angular domain channels and $\bm{B}\in\mathbb{C}^{N \times N}$ is a shift-version discrete Fourier transform (DFT) matrix, whose $n$-th column is given by
\begin{equation}
\bm{b}_{n}=\frac{1}{\sqrt{N}}[1,e^{j\pi \eta_n},\cdots,e^{j\pi \eta_n(N-1)}]^T,
\end{equation}
where $\eta_n = \frac{2n-1}{N}$ for $n=0,1,\cdots,N-1$\cite{VBI}. With the narrow angular spread assumption, the angular domain channels exhibit the spatial-clustered sparsity structure. Specifically, the number of significant elements of each column of $\bm{X}$ is much smaller than $N$ and the significant elements appear in a cluster\cite{VBI}. Besides, for a multi-antenna user, the sparsity structures of all antennas at the user can be considered as same due to their close locations, i.e., the positions of significant elements in all columns of $\bm{X}$ are same, as illustrated in Fig. \ref{system}. By properly exploiting the sparsity structure of the angular domain channels, the estimation performance can be substantially improved, and the estimation overhead can be significantly reduced.  In \cite{VBI}, CS algorithms are used and the sparsity structure is reflected in the prior distribution design of the angular domain channels while we exploit it through DL with the aid of GPS information of the user in this paper.

\subsection{Problem Formulation}
During uplink training, orthogonal pilot sequences are assigned to different antennas at the user. Denote the pilot sequence of the $m$-th antenna as $\bm{p}_m\in\mathbb{C}^{1\times M}$, where the pilot length is set equal to the number of antennas at the user, then the entire pilot matrix can be denoted as $\bm{P}=[\bm{p}_1^T,\bm{p}_2^T,\cdots,\bm{p}_M^T]^T\in \mathbb{C}^{M\times M}$. With the HAD architecture, the signals arriving at antennas at the BS have to go through the phase shifter network first before received by the RF chains. After digital processing, the baseband signal received during the training period can be expressed as\footnote{Since pilot length $M$ is often relatively small, the change of channels during pilot training phase is negligible\cite{VBI}.}
\begin{align}
\bm{Y}=\bm{W}_{BB}\bm{W}_{RF}(\bm{HP}+\bm{N})\in \mathbb{C}^{R \times M},
\label{receive_multi_antennas}
\end{align}
where $\bm{W}_{BB}\in{\mathbb C}^{R\times R}$ and $\bm{W}_{RF}\in{\mathbb C}^{R\times N}$ are the digital and analog combining matrices at the BS, respectively, and $\bm{N}\sim \mathcal{CN}(0,\sigma^2)\in\mathbb{C}^{N\times M}$ is the additive noise at the BS with variance $\sigma^2$. As the phase shifters only change the phase of signals, we have $|[\bm{W}_{RF}]_{i,j}|=1/\sqrt{N}$, $\forall i,j$ after normalization\cite{Structured_SBL}. Without loss of generality, we fix the power of the pilot sequences to unit, i.e., $\bm{p}_i\bm{p}_j^H=0, \forall i \neq j$ and $\bm{p}_i\bm{p}_i^H=1, \forall i$. Specifically, $\bm{P}$ can be set to an $M$-dimensional DFT matrix. Then, the effective signal-to-noise ratio (SNR) can be adjusted by varying the noise variance, i.e., $\text{SNR}=1/\sigma^2$. Exploiting the orthogonality of pilot sequences among different antennas, the baseband signal corresponding to the $m$-th antenna at the user can be obtained by
\begin{equation}
\bm{y}_m = \bm{Y}\bm{p}_m^H = \bm{W}_{BB}\bm{W}_{RF}\bm{h}_m+\widetilde{\bm{n}}_m\in \mathbb{C}^{R \times 1},
\label{receive}
\end{equation}
where $\widetilde{\bm{n}}_m\triangleq \bm{W}_{BB}\bm{W}_{RF}\bm{N}\bm{p}_m^H$ is the effective noise of the $m$-th antenna at the user and ${\bm{h}_m}\in{\mathbb C}^{N\times 1}$ is the uplink channel from the $m$-th antenna at the user to the BS, i.e., the $m$-th column of the channel matrix $\bm{H}$. For brevity, we will consider a specific antenna at the user from now on and omit the subscript $m$. Since $\bm{B}\bm{B}^H=\bm{I}$, (\ref{receive}) can be rewritten in the standard CS form as
\begin{align}
\bm{y}=\bm{\Phi}\bm{x}+\widetilde{\bm{n}},
\label{CS}
\end{align}
where $\bm{x}$ is the angular domain channel of an antenna at the user, i.e., a column of $\bm{X}$, and $\bm{\Phi}\triangleq \bm{W}_{BB}\bm{W}_{RF}\bm{B}\in \mathbb{C}^{R \times N}$ is called the measurement matrix borrowing the term from CS. Once the estimation of $\hat{\bm{x}}$ is obtained based on received signal $\bm{y}$ and measurement matrix $\bm{\Phi}$, the estimation of the original channels can be readily recovered by $\hat{\bm{h}}=\bm{B}\hat{\bm{x}}$.

The conventional method to solve (\ref{CS}) is to use CS algorithms like S-VBI\cite{VBI}, which, however, are not very suitable for the considered problem for the following reasons. First, the true AoAs of uplink channel paths at the BS do not exactly match the discrete angle grids determined by the shift-version DFT matrix, which results in the power leakage, thereby decreasing the sparsity level of $\bm{x}$\cite{BEM3}. In the case of recovering low sparsity level signals, CS algorithms often fail to achieve satisfactory performance. Second, the iterative nonlinear optimization operations involved in CS algorithms incur high computational complexity, which hinders their practical applications, especially in time-sensitive scenarios, such as high mobility communications. Third, the commonly adopted measurement matrix designs in conventional CS algorithms can not effectively exploit the spatial-clustered sparsity structure of the angular domain channels, therefore leading to low signal measurement efficiency.

To circumvent the above issues, algorithms with less dependency on signal sparsity and better performance-complexity tradeoffs are desirable. Motivated by this, we propose a DL-based HAD massive MIMO channel estimation approach. We use DL for channel estimation mainly due to three reasons. First, according to the universal approximation theory, deep neural networks can approach any complex mapping functions between input and output with sufficient training\cite{universal}, therefore is less dependent on the sparsity of angular domain channels. Second, once the network is trained, only multiplication and addition operations of matrices and vectors are required during deployment, which has much lower computational complexity compared with the iterative CS algorithms. Furthermore, through proper design of the network architecture, the measurement matrix, which is consistent with the sparsity structure of angular domain channels, can be jointly learned with the channel estimator, therefore improving the signal measurement efficiency.

\section{DL-based channel estimation framework}
In this section, we first overview the working procedures of the proposed framework, then introduce the key modules in detail, including the angular space segmentation method, how to deal with GPS error, the architecture and layers designs of the network, the training process, the complexity analysis, and the brief discussion about the extension to the multi-user case.

\subsection{Overview of the Framework}
\begin{figure}[htbp]
\centering
\includegraphics[width=1\textwidth]{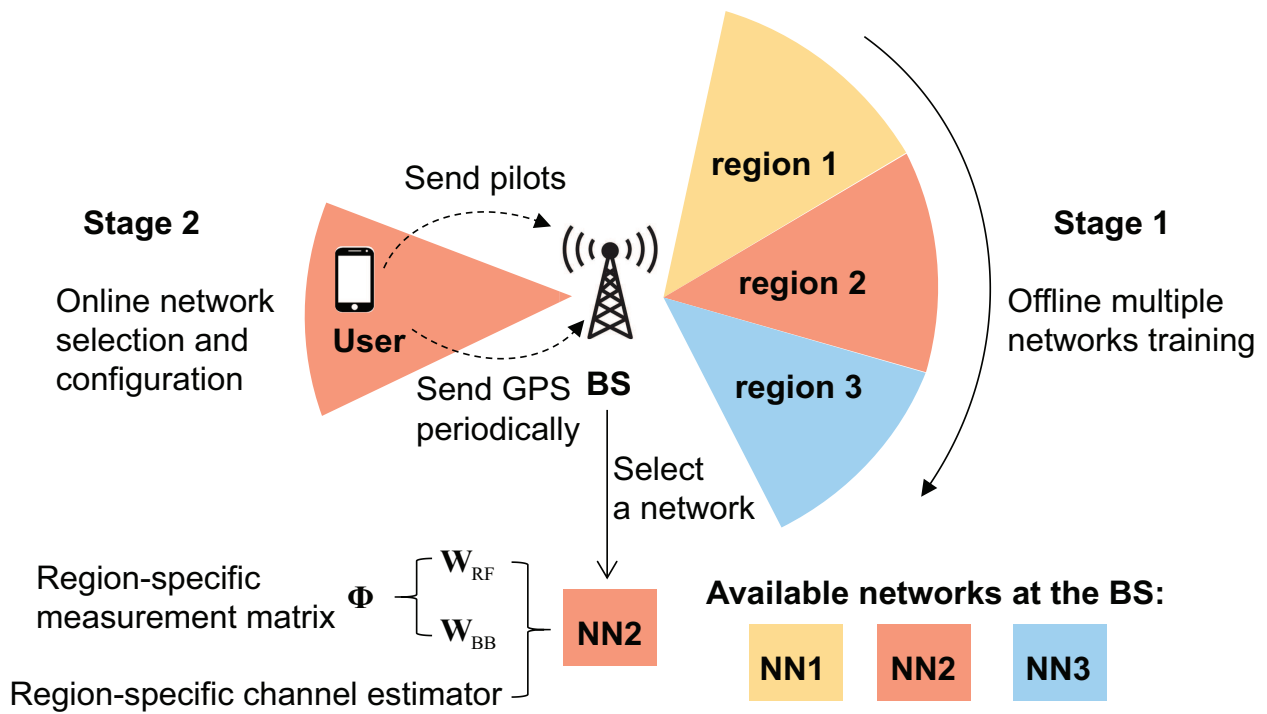}
\caption{Two-stage DL-based channel estimation framework.}
\label{framework}
\end{figure}

Before diving into the details of the proposed approach, let us first briefly introduce the overall framework and the key procedures. As illustrated in Fig. \ref{framework}, the framework contains two stages, namely offline network training, and online network selection and configuration.

Without loss of generality, it is assumed that the users are randomly distributed in the angular space. Due to the spatial-clustered sparsity structure, the angular domain channels of users at close angles relative to the BS have similar features. In contrast, for users at different angles, the channel features also differ significantly. Therefore, instead of training a single neural network for all users at all possible angles, it is more reasonable to train multiple networks for users in different angular regions. Motivated by this, the entire angular space is segmented into many small angular regions in the first stage, then for each region, a dedicated neural network containing measurement matrix and channel estimator is trained with a large amount of channel data collected from users in the region. Then, in the second stage, two kinds of data are sent to the BS by the user. First, the GPS information is sent periodically to help BS select a suitable network for the user. Specifically, based on the received position coordinates, the azimuth angle of the user can be calculated, and the network whose corresponding angular region contains the azimuth angle is selected, from which a pair of region-specific measurement matrix and channel estimator can be extracted. Second, all antennas at the user transmit orthogonal pilot sequences to the BS simultaneously during the pilot training phase. Then, based on the received baseband signal processed by the analog and digital combining matrices, the channel estimator can be used to estimate the user's channels.

Notice that, in this paper, we assume GPS information is available at the user and can be sent to the BS without error through a dedicated link. The selection of network changes slowly even in high mobility scenarios due to the width of segmented angular regions, therefore the frequency of GPS acquisition can be low, e.g., once per dozens of channel coherence blocks, and the overhead is acceptable. Besides, when GPS is not available, the azimuth angle of the user can actually also be estimated based on past channels with the adopted narrow angular spread channel model.

\subsection{Angular Space Segmentation}
To facilitate the multiple networks training and network selection procedures, proper angular space segmentation is of paramount importance. The segmentation granularity entails a fundamental tradeoff between the network performance and training overhead. With more fine-grained segmentation, the performance of individual networks will improve since the learned measurement matrix can be more targeted for channels in a smaller angular region and the learned channel estimator can have better performance with a simpler input distribution. At the same time, however, more networks need to be trained and stored at the BS, which consume more computational and storage resources. In contrast, fewer networks are needed with more coarse-grained segmentation, at the cost of decreased individual network performance.

In principle, the entire angular space needs to be segmented. However, by exploiting the implicit symmetry in the angular domain, only 1/4 angular space needs to be considered due to the following key observations:
\begin{itemize}
\item For a ULA, two angles with the same sine value correspond to the same steering vector. Since $\pi/2$ and $-\pi/2$ are the symmetry axes of the sinusoidal function, only the angular space $[-\pi/2,\pi/2]$ needs to be considered during training, then angular spaces $[-\pi,-\pi/2]$ and $[\pi/2,\pi]$ can be handled by the networks trained for $[-\pi/2,0]$ and $[0,\pi/2]$, respectively.
\item Networks trained for angular space $[0,\pi/2]$ can be used for angular space $[-\pi/2,0]$ with proper modification of the measurement matrix and the input of the channel estimator. For instance, consider User 1 with $\theta^1_i \in[0,\pi/2]$ and User 2 with $\theta^2_i=-\theta^1_i$, $i=1,\cdots,N_p$. Denote the measurement matrix for User 1 as $\bm{\Phi}^1$, then the baseband signal of User 1 is
    \begin{align}
    {\bm y}^1 = \bm{\Phi}^1\bm{x}^1+\widetilde{\bm{n}}^1=\bm{\Phi}^1\bm{B}^H\bm{h}^1+\widetilde{\bm{n}}^1,
    \end{align}
    where $\bm{h}^1$ denotes the channel of User 1. Now, let the measurement matrix for User 2 being $\bm{\Phi}^2=(\bm{\Phi}^1\bm{B}^H)^*\bm{B}$, i.e., $\bm{W}_{BB}^2=(\bm{W}_{BB}^1)^*$ and $\bm{W}_{RF}^2=(\bm{W}_{RF}^1)^*$, then the conjugate of the baseband signal of User 2 can be expressed as
    \begin{align}
    \left({\bm y}^2\right)^* =  \left(\bm{\Phi}^2\bm{x}^2+\widetilde{\bm{n}}^2\right)^*=\left((\bm{\Phi}^1\bm{B}^H)^*\bm{B}\bm{B}^H\bm{h}^2\right)^*+(\widetilde{\bm{n}}^2)^*=\bm{\Phi}^1\bm{B}^H\left(\bm{h}^2\right)^*+(\widetilde{\bm{n}}^2)^*,
    \end{align}
    where $\bm{h}^2$ denotes the channel of User 2. Since the steering vectors of $\bm{h}^1$ and $\left(\bm{h}^2\right)^*$ are the same, it is easy to show that the statistical distributions of ${\bm y}^1$ and $\left({\bm y}^2\right)^*$ are the same. Therefore, the two users can share the same channel estimator network. Finally, given the network output as $\hat{\bm{x}}^1$, the angular domain channel prediction of User 2 can be obtained by
    \begin{align}
    \hat{\bm{x}}^2=\bm{B}^H(\bm{B}\hat{\bm{x}}^1)^*.
    \end{align}
\end{itemize}

As confirmed by experimental results, the performance obtained by the above methods is exactly the same as segmenting and training networks from scratch with true angles throughout the entire angular space, validating their correctness. 

Now, denote the width of each angular region as $\beta$, the number of networks need to be trained and stored at the BS is only $N_\text{net}=\lceil\frac{\pi/2}{\beta}\rceil$.

\subsection{Dealing with GPS Error}
In practice, the GPS positioning error needs to be considered, otherwise it will lead to performance degradation for users near the edges of segmented angular regions. It is possible that the network trained for the adjacent region is mistakenly selected due to GPS error, leading to the usage of measurement matrix and channel estimator mismatching the current channel distribution. To solve this issue, we expand the angular ranges of training data. When the GPS information is accurate, for a certain angular region, only channels of users whose azimuth angles are inside this region need to be collected to construct the dataset for the training of the dedicated network. Specifically, the angular range of the training data is $[\theta_{start},\theta_{end}]$ for angular region $[\theta_{start},\theta_{end}]$, where $\theta_{start}$ and $\theta_{end}$ denote the start and end azimuth angles of the region, respectively. However, if the maximal azimuth angle error caused by the GPS positioning error is $\bigtriangleup\theta_{az}$, then the range should be set to $[\theta_{start}-\bigtriangleup\theta_{az},\theta_{end}+\bigtriangleup\theta_{az}]$, which means that the ranges of two adjacent regions are overlapped by $\bigtriangleup\theta_{az}$. By expanding the angular range of the training data, even if the network of the adjacent region is selected by mistake, the measurement matrix and channel estimator can still match the channel distribution since it has been already considered during the training process. The performance degradation only comes from the wider angular range of the training data, which is marginal since $\bigtriangleup\theta_{az}$ is relatively small due to the long distance between the BS and the user.

Overall, the comprehensive and detailed processes are summarized in Algorithm 1.
\begin{algorithm}[htbp]
  \caption{The offline networks training, online network selection and configuration processes}
  \begin{algorithmic}[1]
    \State {\bf $\%$ Initialization}
	\State Initialize the width of angular regions $\beta$, the maximal azimuth angle error $\bigtriangleup\theta_{az}$, the angular spread of uplink channel $\bigtriangleup\theta$, dataset $\mathcal{D}$, required data number $D$, data number counter $d=0$, the start angle $\theta_{start}=0$, and the lower bound of the end angle $\theta_{elb}=\pi/2$
    \State {\bf $\%$ Offline multiple networks training }
	\State {\bf while} $\theta_{s}<\theta_{elb}$ {\bf do}
	\State{\bf while} $d<D$ {\bf do}
	\State Compute the end angle $\theta_{end}=\theta_{start}+\beta$, uniformly sample an azimuth angle of the user $\theta_{az}$ in the angular region $[\theta_{start}-\bigtriangleup\theta_{az},\theta_{end}+\bigtriangleup\theta_{az}]$
	\State Uniformly sample AoAs of channel paths at the BS in the AoA range $[\theta_{az}-\bigtriangleup_{\theta},\theta_{az}+\bigtriangleup_{\theta}]$, uniformly sample AoDs of channel paths at the user in the AoD range $[0,2\pi]$, randomly generate gains of channel paths according to $\mathcal{CN}(0,1)$. Then, generate a channel label $\bm{H}_d$ according to (\ref{channel}) and obtain the data $\hat{\bm{H}}_d$ by adding noise on the label
	\State Append $\mathcal{D}$ with $M$ ($\hat{\bm{h}}_d$, $\bm{h}_d$) pairs, where $\hat{\bm{h}}_d$ and $\bm{h}_d$ are columns of $\hat{\bm{H}}_d$ and $\bm{H}_d$, respectively
	\State $d=d+1$
	\State{\bf end while}
	\State Train a neural network to convergence with $\mathcal{D}$, save the network model and record the $\theta_{start}$ and $\theta_{end}$ corresponding to this network. Then, empty $\mathcal{D}$
	\State Update $\theta_{start}=\theta_{end}$ for the next angular region
    \State{\bf end while}
    \State {\bf $\%$ Online network selection}
	\State Compute the azimuth angle of the user $\theta_{az}$ based on the position coordinates of the BS and the user, which can be obtained by GPS information
	\State If $0\le\theta_{az}\le\pi/2$, select the network that satisfies $\theta_{start}\le\theta_{az}\le\theta_{end}$. If $\pi/2\le\theta_{az}\le\pi$, select the network corresponding to the azimuth angle with the same sine value. If $\pi\le\theta_{az}\le2\pi$, select the network corresponding to the azimuth angle with the opposite sine value
	\State {\bf $\%$ Measurement matrix configuration and channel estimation}
    \State Extract the analog and digital measurement matrices $\bm{W}_{RF}$ and $\bm{W}_{BB}$, and the channel estimator from the selected network
    \State Configure the phase shifters according to $\bm{W}_{RF}$
	\State Measure the signals and process the baseband signal with $\bm{W}_{BB}$, then do channel estimation with the channel estimator
    \label{Algo_segmentation}
  \end{algorithmic}
\end{algorithm}

\subsection{Joint Learning Network Design Based on the Autoencoder Architecture}
\begin{figure}[htbp]
\centering
\includegraphics[width=1\textwidth]{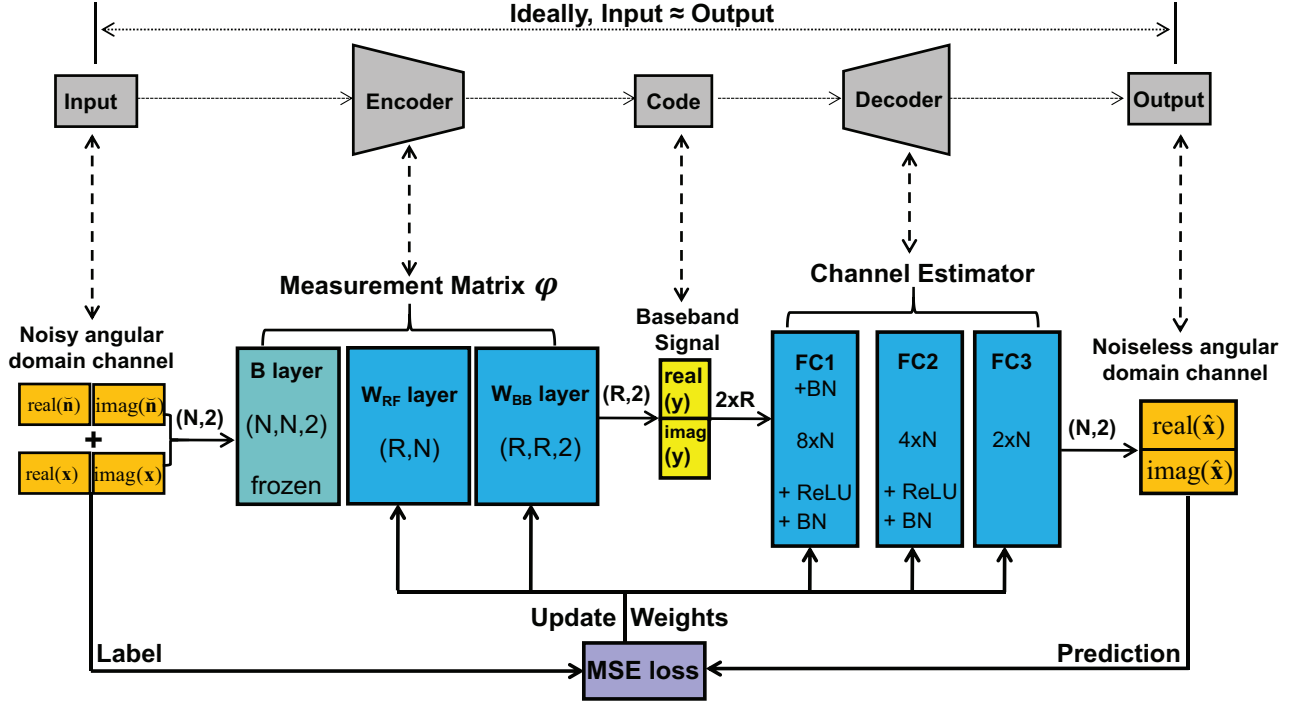}
\caption{Autoencoder for joint learning of the measurement matrix and the channel estimator.}
\label{AE}
\end{figure}

Inspired by \cite{DL_CE2,DL_CE3,Beamforming3}, in this paper, the joint learning of the measurement matrix and the channel estimator is achieved by autoencoder. As illustrated in the upper part of Fig. \ref{AE}, in the original autoencoder, the input first goes through an encoder to get the low-dimensional code and then the decoder tries to recover the input at the output based on the code. If the input is sparse on certain basis and the dimension of the code is sufficient relative to the sparsity level, the input can be perfectly recovered with sufficient training. It is very lucky that the considered problem can also be modeled in the way of autoencoder. Specifically, as illustrated in Fig. \ref{AE}, with angular domain channels $\bm{x}$ being the input, the signal measurement process can be regarded as the encoder, then baseband signal $\bm{y}$ is obtained as the code and the channel estimator can be regarded as the decoder. To simplify the implementation, the noise is added to the angular domain channels directly\cite{Beamforming2}. Rewrite (\ref{CS}) as $\bm{y}=\bm{\Phi}(\bm{x}+\check{\bm{n}})$, where $\check{\bm{n}}\triangleq \bm{B}^H\bm{Np}^H$ is the equivalent angular domain channel noise, then the input of autoencoder becomes the noisy angular domain channels $\bm{x}+\check{\bm{n}}$, and the noiseless angular domain channels $\bm{x}$ are recovered at the output. The mean-squared error (MSE) between the input and output is used as the loss function, which can be expressed as
\begin{equation}
\text{MSE loss}=\frac{1}{n}\sum_{i=1}^{n}\left\|\hat{\bm{x}}_i-\bm{x}_i\right\|^2,
\end{equation}
where $n$ is the number of data samples in a mini-batch. The loss reflects the effectiveness of the current measurement matrix and the channel estimator and guides the learning process of the neural network. According to the loss and the back propagation rule, the gradients of weights in each layer can be computed and the weights can be updated based on the gradients and the learning rate. Notice that, the real and imaginary parts of variables have to be separately processed in the network since complex training is still not well supported by current DL libraries.

Notice that, similar joint learning methods realized by autoencoders have already been used in massive MIMO channel estimation\cite{DL_CE2,DL_CE3,Beamforming3}. Nevertheless, beamformers are directly predicted in \cite{Beamforming3} and pilot matrices are optimized in \cite{DL_CE2,DL_CE3}. In this paper, we innovatively propose to change measurement matrices to match channel distributions of users in different positions with the aid of GPS information. 

\subsection{Network Layers}
For the measurement matrix learning, there are two kinds of feasible designs, one is to separately optimize $\bm{W}_{BB}$ and $\bm{W}_{RF}$, and the other is to directly optimize $\bm{\Phi}$ and solve $\bm{W}_{BB}$ and $\bm{W}_{RF}$ that satisfy $\bm{W}_{BB}\bm{W}_{RF}\bm{B}=\bm{\Phi}$. The latter design can realize the fully digital processing and have better performance while twice the number of phase shifters have to be used in order to ensure the existence of feasible $\bm{W}_{BB}$ and $\bm{W}_{RF}$ solutions\cite{prove_fully_digital}. Considering the low cost of phase shifters compared with RF chains, it can be an option when high performance is the top priority. In this paper, we use the former design, which is consistent with regular HAD systems.

As illustrated in Fig. \ref{AE}, we customize three consecutive network layers, namely ``$\bm{B}$ layer", ``$\bm{W}_{RF}$ layer", and ``$\bm{W}_{BB}$ layer", to optimize the measurement matrix $\bm{\Phi}$. Each of the three layers realizes the function of a left multiplication of the previous layer's output by a matrix defined by the layer's weights. Besides, the multiplication of complex matrices is realized by modifying the forward computation rules of the layer, where the real and imaginary parts are computed separately\footnote{We use the TensorFlow DL library, which supports the customization of network layers where the forward computation rules can be self-defined and the computation and backpropagation of gradients are realized automatically.}. Specifically, the ``$\bm{B}$ layer" contains a group of $(N,N,2)$-dimensional weights that represent the elements of matrix $\bm{B}$. The ``$\bm{W}_{RF}$ layer" contains a group of $(R,N)$-dimensional weights that represent the phases of phase shifters, then the real and imaginary parts can be obtained by computing the sine and cosine values of these phases. The ``$\bm{W}_{BB}$ layer" contains a group of $(R,R,2)$-dimensional weights that represent the elements of matrix $\bm{W}_{BB}$. The weights of the ``$\bm{B}$ layer" are frozen and those of the other two layers are trainable. Consequently, the real and imaginary parts of baseband signal $\bm{y}$ can be obtained at the output of the ``$\bm{W}_{BB}$" layer. After training, the weights of the ``$\bm{W}_{RF}$ layer" and the ``$\bm{W}_{BB}$ layer" can be fetched out to form the optimized analog and digital combining matrices $\bm{W}_{RF}$ and $\bm{W}_{BB}$, respectively.

For the channel estimator in \cite{DL_CE2}, the so called deep unfolding method is used, where the iterations of conventional CS algorithms are unfolded into the cascade of network layers and algorithm parameters like step size are trainable to improve the estimation performance. While for the considered problem, deep unfolding is not a wise choice. On the one hand, the angular domain channel is not sparse enough due to power leakage and CS algorithms cannot achieve satisfactory performance in such case. To seek remarkable performance breakthrough, the architecture of CS algorithms should be abandoned. On the other hand, the computational complexity of the unfolded neural networks is even higher than the original CS algorithms. So, we directly learn the mapping function of channel estimation with a deep neural network in a data-driven manner.

\begin{figure}[!htb]
\centering
\includegraphics[width=0.8\textwidth]{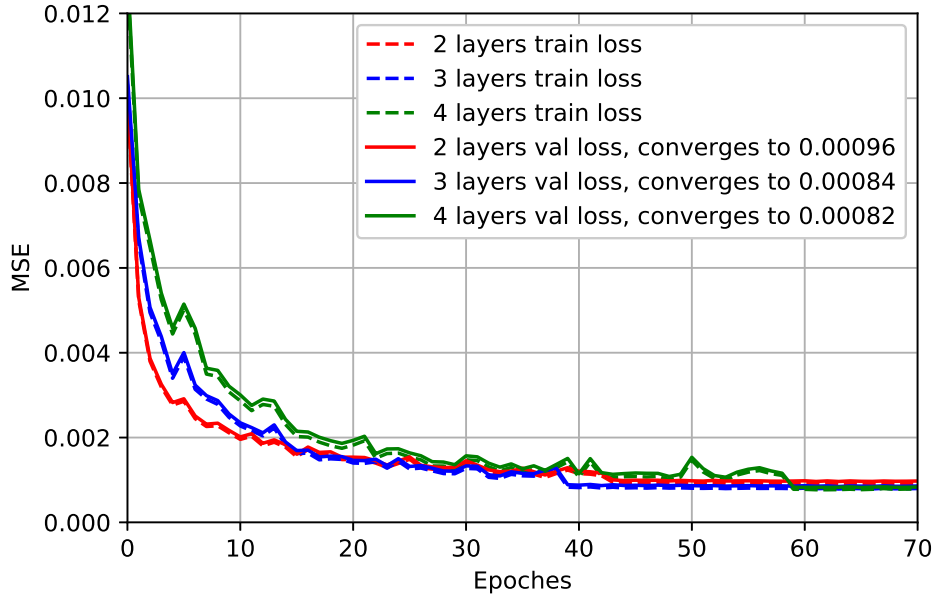}
\caption{The impact of network depth on loss convergence.}
\label{impact_of_depth}
\end{figure}

There are many network architecture candidates, including fully-connected neural network (FNN), CNN, long short term memory network, etc. We use FNN since there exists little local correlation in received signal $\bm{y}$ and we do not consider time series data here. Specifically, we use three fully-connected (FC) layers with $8N$, $4N$, and $2N$ neurons, respectively. The choices of the neuron numbers and the network depth are determined by cross validation. For instance, when $N=64,R=16,\text{SNR}=20$ dB, the impact of network depth is illustrated in Fig. \ref{impact_of_depth}. As we can see, the adopted network architecture with three FC layers achieves the best performance-complexity tradeoff. Besides, a batch normalization (BN)\cite{BN} layer is inserted before the first FC layer to preprocess the features fed to the channel estimator, as well as between each two FC layers to avoid gradient vanishing or explosion and promote efficient training. Except for the output FC layer, which does not need activation, the activation function for the other two FC layers is rectified linear unit (ReLU).

\subsection{Training Process of the Network}
As for network training, the weights of the encoder is initialized by the conventional measurement matrix, which will be detailed later. The decoder uses the widely adopted Xavier initializer\cite{xavier}. The Adam optimizer\cite{adam} is used to update weights and the initial learning rate and batch size are set to $0.001$ and $128$, respectively. To balance the training complexity and testing performance, we generate $50,000$ channels according to the channel model as labels and add noise according to the transmission model to obtain the corresponding data. Notice that, the total number of samples is actually $50,000M$ since the multi-antenna channel estimation problem is decomposed into multiple single-antenna problems. Then, the whole dataset is split into a training set and a validation set with a ratio of $4:1$. In order to accelerate convergence at the beginning and reduce loss oscillation near the end of training process, learning rate decays with a factor of $0.1$ when the validation loss does not decrease for $5$ consecutive epochs. Besides, to prevent overfitting and obtain the model with best validation performance, the training process is terminated when the validation loss does not decrease for $12$ consecutive epochs, i.e., early stopping\cite{early_stopping}.

\subsection{Complexity Analysis}
Since the networks are trained offline and the BS usually has powerful computing ability, we care more about the online prediction complexity. Notice that, multiple offline-trained networks will not increase the online complexity since only one of the networks is selected for signal measurement and channel estimation. For a FC layer, the complexity is $\mathcal{O}(N_{in}N_{out})$ where $N_{in}$ and $N_{out}$ denote the neuron numbers of the previous layer and the current layer, respectively. Therefore, for the designed channel estimator part, the complexity of estimating channels of all antennas of a user can be readily computed as $\mathcal{O}(M(16RN+40N^2))$. For the main CS baseline S-VBI, the complexity is approximately $\mathcal{O}(I_EM(\frac{2}{3}R^3+(2R+\frac{44M-49}{M}+7)N^2))$, where $I_E$ denotes the number of iterations\cite{VBI}. As we can see, if a fixed radio frequency chain ratio $R/N$ is considered, then the complexity of the proposed network only scales linearly with $M$ and squarely with $N$, while the complexity of S-VBI scales linearly with $M$ and cubically with $N$. Besides, the network does not require multiple iterations either. In typical system settings, the complexity of S-VBI is dozens of times higher than neural network, and the advantage of neural network in real running time can be even more exaggerated if GPU is used for testing acceleration. The low complexity further strengthens the practical value of the proposed approach, especially in scenarios, like high mobility communications, where the channel is fast fading and the estimation process needs to be completed very quickly.

\subsection{Extension to the Multi-User Case}
In the multi-user case, since the analog combining matrix realized by phase shifters has to be fixed during the channel estimation process of a certain user's channels, there are two optional estimation schemes, one is estimating different users' channels in turn with different pairs of matching measurement matrices and channel estimators, the other is estimating all users' channels simultaneously with a fixed pair of measurement matrix and channel estimator that can be trained with channels from the entire angular space. 

In the context of fixed pilot symbol power, there exists a fundamental tradeoff that the former scheme can enjoy the gain brought by the matching measurement matrix and channel estimator while the latter scheme has higher effective SNR thanks to larger pilot sequence energy. In general, estimating in turn can be superior when the number of users $K$ is small since the gain overwhelms the loss in effective SNR, while when $K$ is larger than a certain turning point, estimating simultaneously should be adopted. In practice, to increase the turning point of $K$ and expand the scope of applicable scenarios of the proposed approach, user grouping can be performed since channels of users in the same angular region can still be estimated simultaneously.

In the context of fixed pilot sequence energy, however, the effective SNRs of the above two schemes are the same. Therefore, the proposed approach can be particularly appealing in scenarios where energy saving is critical, such as IoT communications, since much better performance can be achieved without any extra energy cost of channel estimation.

\section{Simulation Results}
\label{simulation}
In this section, extensive simulation results are presented to demonstrate the performance of the proposed DL-based channel estimation approach and validate its superiority. We use the normalized MSE (NMSE) as the performance metric, which is defined by
\begin{equation}
\text{NMSE}=\frac{\left\| \hat{\bm{H}}-\bm{H} \right\|^2}{\left\| \bm{H} \right\|^2},
\end{equation}
where $\hat{\bm{H}}$ denotes the estimated channel matrix. Notice that, transforming channels to the angular domain does not change the $l$-2 norm since $\bm{B}$ is a unitary matrix. Therefore, the NMSEs of the angular domain channels and the original channels are the same. Unless otherwise specified, the default parameters used in the simulation experiments are summarized in Table \ref{parameters}. Besides, all the following results are obtained by averaging $1,000$ randomly generated testing samples.
\begin{table}[htbp]
\centering
\begin{tabular}{|c|c|}
\hline
Parameter & Value \\
\hline
$N$ & $64$\\
\hline
$M$ & $4$\\
\hline
$R$ & $16$\\
\hline
$N_p$ & $20$\\
\hline
$\bigtriangleup\theta$ & $5^\circ$\\
\hline
$\bigtriangleup\theta_{az}$ & $1^\circ$\\
\hline
$\beta$ & $5^\circ$\\
\hline
SNR & $20$ dB\\
\hline
\end{tabular}
\caption{Default simulation parameters.}
\label{parameters}
\end{table}

%\begin{table}[htbp]
%\centering
%\begin{tabular}{|c|c|}
%\hline
%Parameter & Value \\
%\hline
%Initializer & Xavier \\
%\hline
%Optimizer & Adam\\
%\hline
%Init Lr & 0.001\\
%\hline
%Data Number & $50,000$\\
%\hline
%Data split & Training:Validation=4:1\\
%\hline
%Strategy & Lr decay $\&$ Early stopping\\
%\hline
%\end{tabular}
%\caption{Default simulation parameters.}
%\label{parameters}
%\end{table}

\subsection{DL-based Channel Estimator and Learned Measurement Matrix}
For performance comparison, we select three CS algorithms, namely SBL\cite{HAD_CS2}, OMP\cite{OMP}, and the advanced S-VBI proposed in \cite{VBI}, as baselines\footnote{The sparsity corresponding to best performance is searched for OMP. SBL uses the open source code on $\text{https://github.com/pillowlab/DRD/blob/master/ARD/SBL}$. The source code of S-VBI is provided by authors of \cite{VBI} and the common sparsity structures of antennas at the user are exploited to further improve estimation performance.}. For the conventional measurement matrix, we assign a matrix consisting of length-$N$ Zadoff-Chu sequences with different shifting steps\cite{LS,Structured_SBL} to $\bm{W}_{RF}$, and a $R$-dimensional identity matrix to $\bm{W}_{BB}$. Due to space limitation, only $[10^\circ,15^\circ]$ is selected as the representative angular region here.

As can be seen from Table \ref{impact_of_joint_training}, with either the conventional or the learned measurement matrix, the proposed approach outperforms all CS algorithms significantly with different RF chain ratios $R/N$, which demonstrates the superiority of the DL-based channel estimator. On the other hand, for most CS algorithms and the DL-based channel estimator, the jointly learned measurement matrix can significantly improve the channel estimation performance compared with the conventional one, which demonstrates the benefits of the measurement matrix learning. According to the initial simulation results above and in \cite{VBI}, the performance of S-VBI is much better than SBL and OMP, hence only S-VBI is used for comparison in the following experiments for brevity.

To further show the superiority of the proposed approach, we also compare with LS estimation, where only channels of $R$ antennas are estimated at once due to a limited number of RF chains and the pilot overhead is $N/R$ times more than CS algorithms and DL-based approach. We can see from Table \ref{impact_of_joint_training} that with 1/4 RF chains, the performance of S-VBI with the conventional measurement matrix is worse than LS and it surpasses LS with the learned one, while the DL-based approach outperforms LS significantly with both the conventional and learned measurement matrices. With 1/8 RF chains, S-VBI is inferior to LS with both the conventional and learned measurement matrices while DL still performs better than LS with the learned measurement matrix.
\begin{table}[htbp]
\centering
\begin{tabular}{|c|c|c|c|c|}
\hline
\multirow{2}{*}{Algorithms} & \multicolumn{2}{c|}{Conventional $\bm{\Phi}$} & %
    \multicolumn{2}{c|}{Learned $\bm{\Phi}$}\\
\cline{2-5}
 & 1/4 RF chains & 1/8 RF chains & 1/4 RF chains & 1/8 RF chains \\
\hline
SBL & 0.36643 & 1.2918 & 0.16626 & -\\
\hline
OMP & 0.63112 & 1.23457 & 0.07258 & 0.07362\\
\hline
S-VBI & 0.01345 & 0.90671 & 0.00461 & 0.02713\\
\hline
LS & 0.01070 & 0.01070 & 0.01070 & 0.01070\\
\hline
\bf{DL} & $\bm{0.00662}$ & $\bm{0.02783}$ & $\bm{0.00172}$ & $\bm{0.00479}$\\
\hline
\end{tabular}
\caption{NMSE performance of different algorithms in angular region $[10^\circ,15^\circ]$. DL with the conventional $\bm{\Phi}$ is obtained by freezing ``$\bm{W}_{BB}$" and ``$\bm{W}_{RF}$" layers and only train the channel estimator part. Besides, SBL fails with the learned $\bm{\Phi}$ with 1/8 RF chains and the NMSE is large.}
\label{impact_of_joint_training}
\end{table}

\begin{figure}[htbp]
\centering
\includegraphics[width=0.8\textwidth]{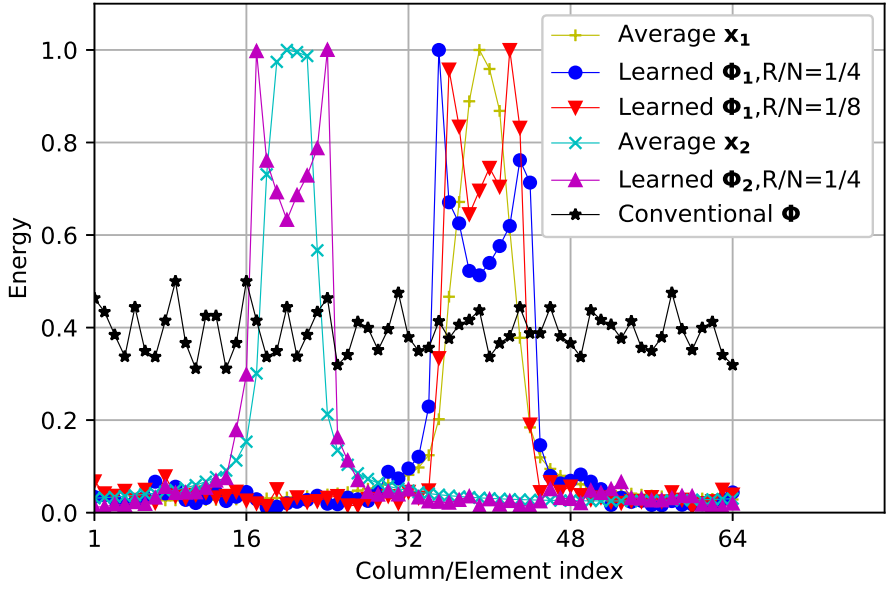}
\caption{Energy distributions of different $\bm{\Phi}$'s columns and the average of $\bm{x}$'s elements. Subscript 1 and 2 denote angular regions $[10^\circ,15^\circ]$ and $[-25^\circ,-20^\circ]$, respectively. Since the angular domain channels and equivalent noise are measured by the same matrix, we perform proper magnitude normalization for better display effects.}
\label{energy}
\end{figure}

To have a more intuitive understanding of the properties of the learned measurement matrix, the energy distributions of its columns and the elements of the average angular domain channel vector are illustrated in Fig. \ref{energy}. As can be seen, for two plotted different angular regions, most columns of the learned measurement matrices are both near zero except for a narrow ``attention zone" where the average angular domain channels are significant. Besides, due to power leakage and the adopted uniformly distributed AoAs inside the angular spread, the average angular domain channels show a peak shape. Correspondingly, for a certain angular region, the energy distribution of the columns of the learned measurement matrix exhibits a bowl shape, where the energy for the peak is low since the average channel energy of these elements are already big enough to be accurately estimated while those elements with lower average channel energy are allocated more measurement energy to minimize the overall estimation error. In contrast, all columns of the fixed conventional measurement matrix have similar energies, and the signal measurement efficiency is much lower without the effective exploitation of the sparsity structure of angular domain channels. The impact of RF chain ratio is also illustrated that, the ``attention zone" is wider and the energy is more dispersed throughout the columns with more RF chains for a certain angular region, because more angles can be taken into account simultaneously with higher degree of freedom. 

\begin{figure}[htbp]
\centering
\includegraphics[width=0.8\textwidth]{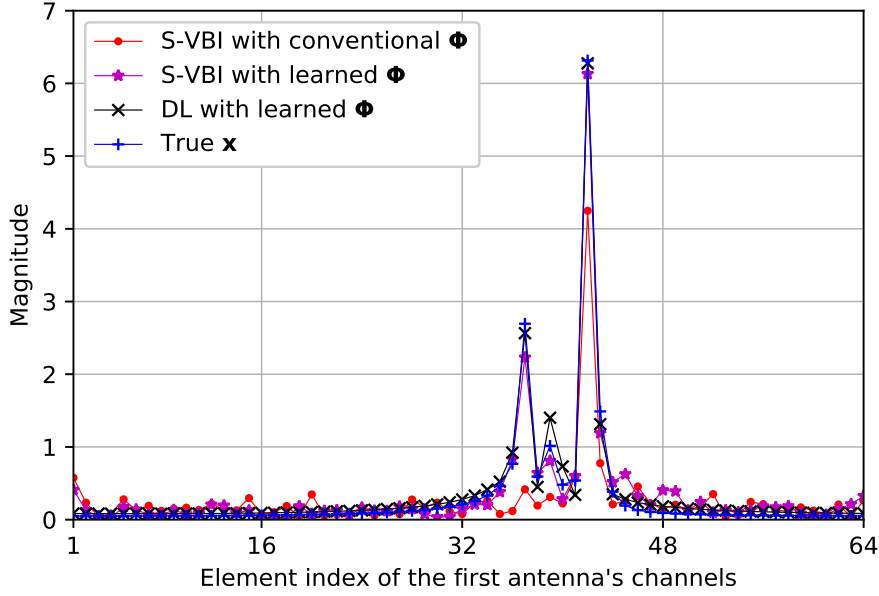}
\caption{A concrete example of estimations made by different algorithms. The NMSEs of the red, magenta, and black curves are 0.2752, 0.0554, and 0.0262, respectively.}
\label{example}
\end{figure}

Fig. \ref{example} shows an estimation example when $N=64$ and $R=8$. As can be observed, S-VBI with the conventional measurement matrix has the worst estimation results and the mismatch between the significant elements is quiet large. With the usage of the learned measurement matrix, the mismatch decreases and the NMSE performance improves. However, there are several incorrect estimation points outside the significant elements region on both S-VBI curves. On the contrary, this phenomenon does not exist on the curve of DL. We also notice that the differences between $\bm{y}$ and $\bm{\Phi\hat{x}}$ for three methods are all small, so it can be concluded that the proposed approach is able to focus on the dedicated angular range and exclude confusing estimations which have small errors in the received signal but are actually impossible for the current angular region.

\subsection{Impact of System Parameters}
\begin{figure}[htbp]
\centering
\includegraphics[width=0.8\textwidth]{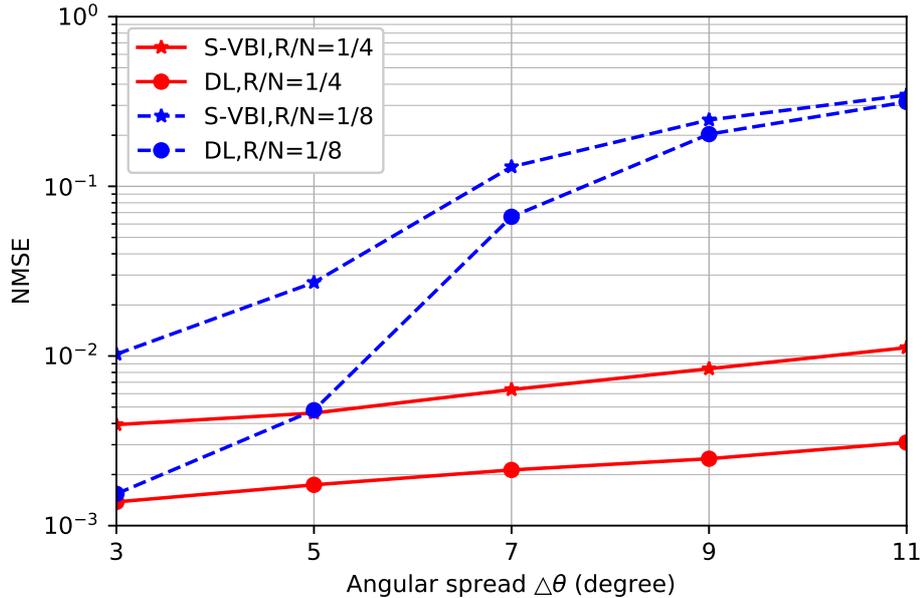}
\caption{The impact of angular spread.}
\label{impact_of_AS}
\end{figure}

In this subsection, we investigate the impacts of various key system parameters. First of all, the impact of angular spread, $\bigtriangleup\theta$, is illustrated in Fig. \ref{impact_of_AS}. As can be seen, the NMSE increases with angular spread in general. It is mainly because the angular domain channels become less sparse with larger angular spread and makes the channel estimation problem more difficult. Besides, the distribution of channels in an angular region becomes more diverse, which leads to a wider ``attention zone" of the learned measurement matrix and lower signal measurement efficiency for a particular channel sample. We also notice that the performance degradation is more severe with 1/8 RF chains, which is caused by insufficient degree of freedom compared with the sparsity of angular domain channels. In contrast, when 1/4 RF chains are used, the degree of freedom is somewhat saturated so that the performance degradation is marginal. Last, when 1/8 RF chains are used, the performance gain of DL compared with S-VBI decreases with the increase of angular spread, indicating that the DL-based channel estimator is better with sparser channels.

Then, the impact of angular space segmentation granularity is illustrated in Table \ref{impact_of_beta}, which is reflected by the width of segmented angular regions $\beta$. As can be seen, smaller $\beta$ results in lower NMSE since the diversity of channel distribution is smaller and the learned measurement matrix is more targeted. Nevertheless, the training complexity and memory consumption are increased with finer-grained segmentation at the same time. Luckily, the overhead of the proposed approach is relatively low, e.g., when $N=64,R=16$, each model only occupies 812 kB memory and takes about 5 minutes to converge on a GTX 1080 Ti GPU. In practice, to strike a balance between training overhead and testing performance, smaller $\beta$ can be used when $\bigtriangleup\theta$ is large so as to compensate for performance degradation at the cost of larger overhead, and larger $\beta$ can be used to reduce overhead at the cost of slightly reduced performance when $\bigtriangleup\theta$ is small.
\begin{table}[htbp]
\centering
\begin{tabular}{ | c | c |  c | c |}%p{2cm}<{\centering}
\hline
$\beta$ & $N_\text{net}$ & NMSE of DL & NMSE of S-VBI\\
\hline
$3^\circ$ & 30 & 0.00165 & 0.00456\\
\hline
$5^\circ$ & 18 & 0.00172 & 0.00461\\
\hline
$10^\circ$ & 9 & 0.00216 & 0.00503\\
\hline
$15^\circ$ & 6 & 0.00251 & 0.00564\\
\hline
\end{tabular}
\caption{The impact of angular space segmentation granularity.}
\label{impact_of_beta}
\end{table}

\begin{figure}[htbp]
\centering
\includegraphics[width=0.8\textwidth]{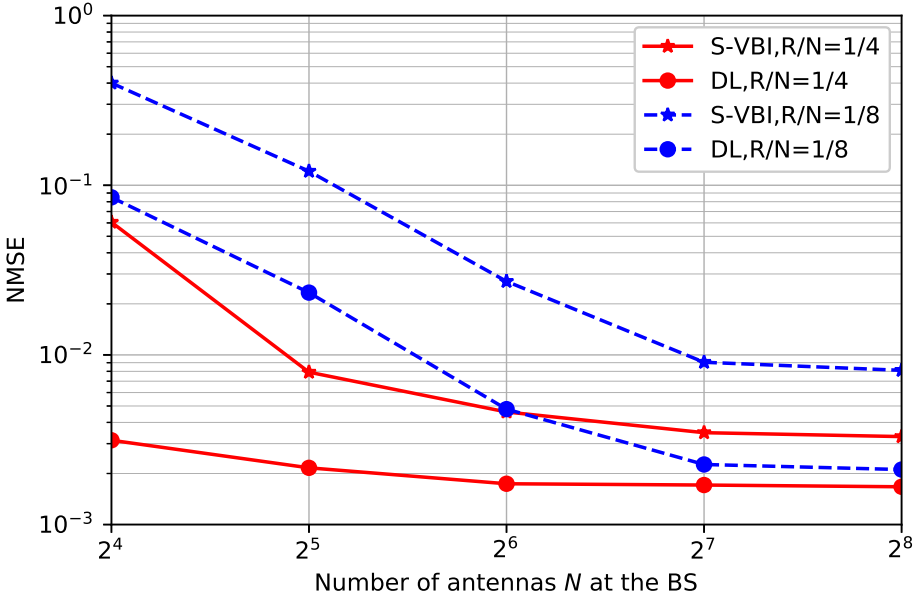}
\caption{The impact of antenna number.}
\label{impact_of_Nt}
\end{figure}

The impact of antenna number at the BS, $N$, is illustrated in Fig. \ref{impact_of_Nt}. In general, with fixed RF chain ratio, the NMSE decreases as $N$ increases because the power leakage is inversely proportional to the antenna number\cite{BEM3}, and the increased sparsity of the angular domain channels brought by more antennas at the BS can simplify the channel estimation problem. Besides, the NMSEs of both algorithms meet floor when $N$ is large enough while the minimal $N$ required to reach the floor increases with the decrease of the RF chain ratio. The reason is that the degree of freedom is easier to saturate with the increase of antenna number with more RF chains. Consequently, when 1/4 RF chains are used, the performance gain of DL compared with S-VBI decreases with the increase of $N$. From the perspective of resource saving, the proposed DL-based approach is also superior. As can be seen from Fig. \ref{impact_of_Nt}, the NMSEs of DL with 1/8 RF chains are lower than those of S-VBI with 1/4 RF chains when $N>64$. On the one hand, if certain NMSE performance is targeted, DL-based approach can save more than half of the hardware and energy consumption of S-VBI. On the other hand, when the RF chain ratio is fixed, DL-based approach can save more than half of the pilot overhead of S-VBI for data transmission, which is particularly appealing when the channel coherence time is short, e.g., in high mobility scenarios. 

\begin{figure}[htbp] %强制位置: H
	\centering  %图片全局居中
	\vspace{-0.35cm} %设置与上面正文的距离
	\subfigtopskip=2pt %设置子图与上面正文或别的内容的距离
	\subfigbottomskip=1pt %设置第二行子图与第一行子图的距离，即下面的头与上面的脚的距离
    \subfigcapskip=-3pt %设置子图与子标题之间的距离
	\subfigure[R/N=1/4]{
		\label{impact_of_snr_4}
		\includegraphics[width=0.485\textwidth]{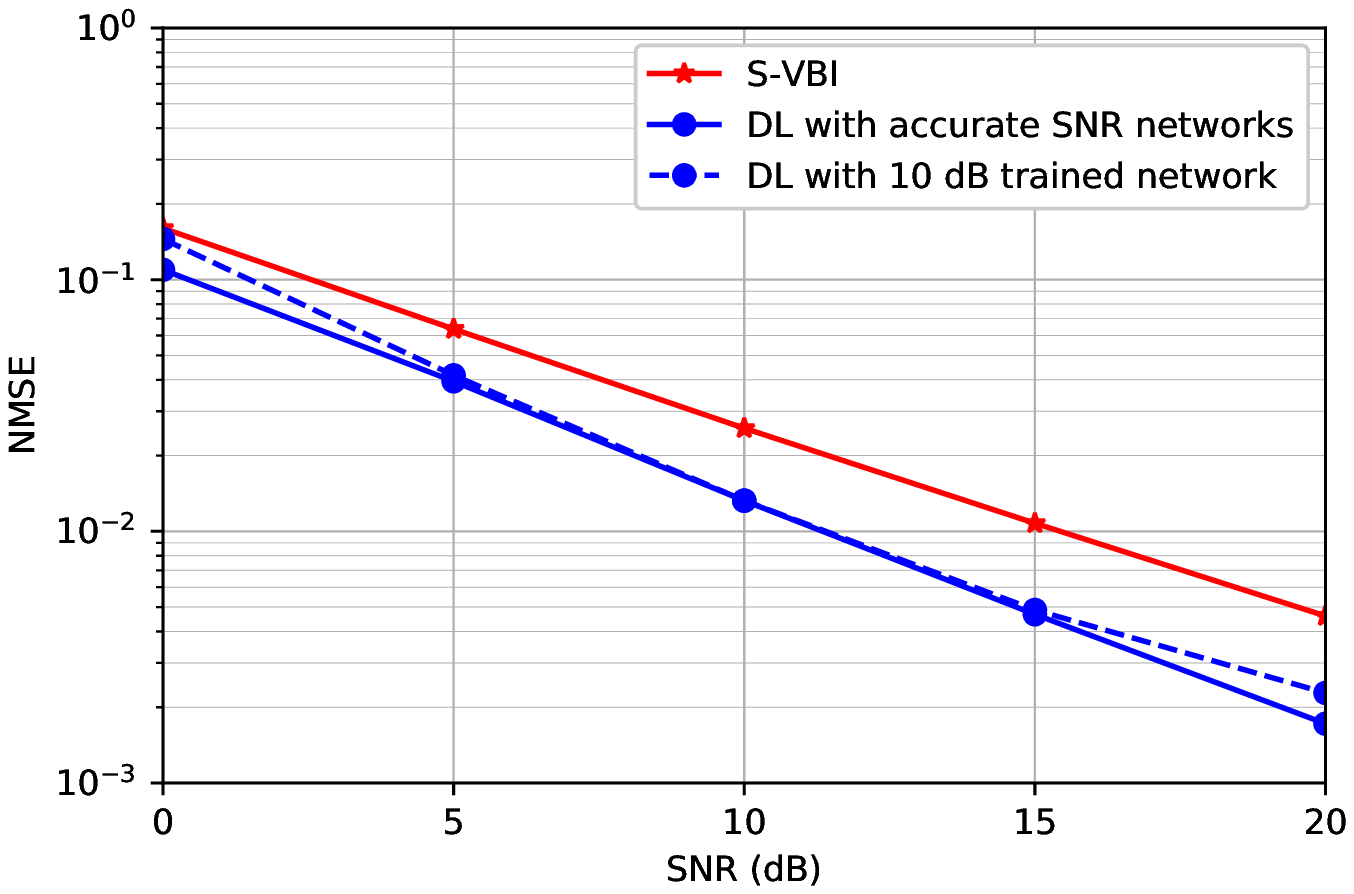}}
	\subfigure[R/N=1/8]{
		\label{impact_of_snr_8}
		\includegraphics[width=0.485\textwidth]{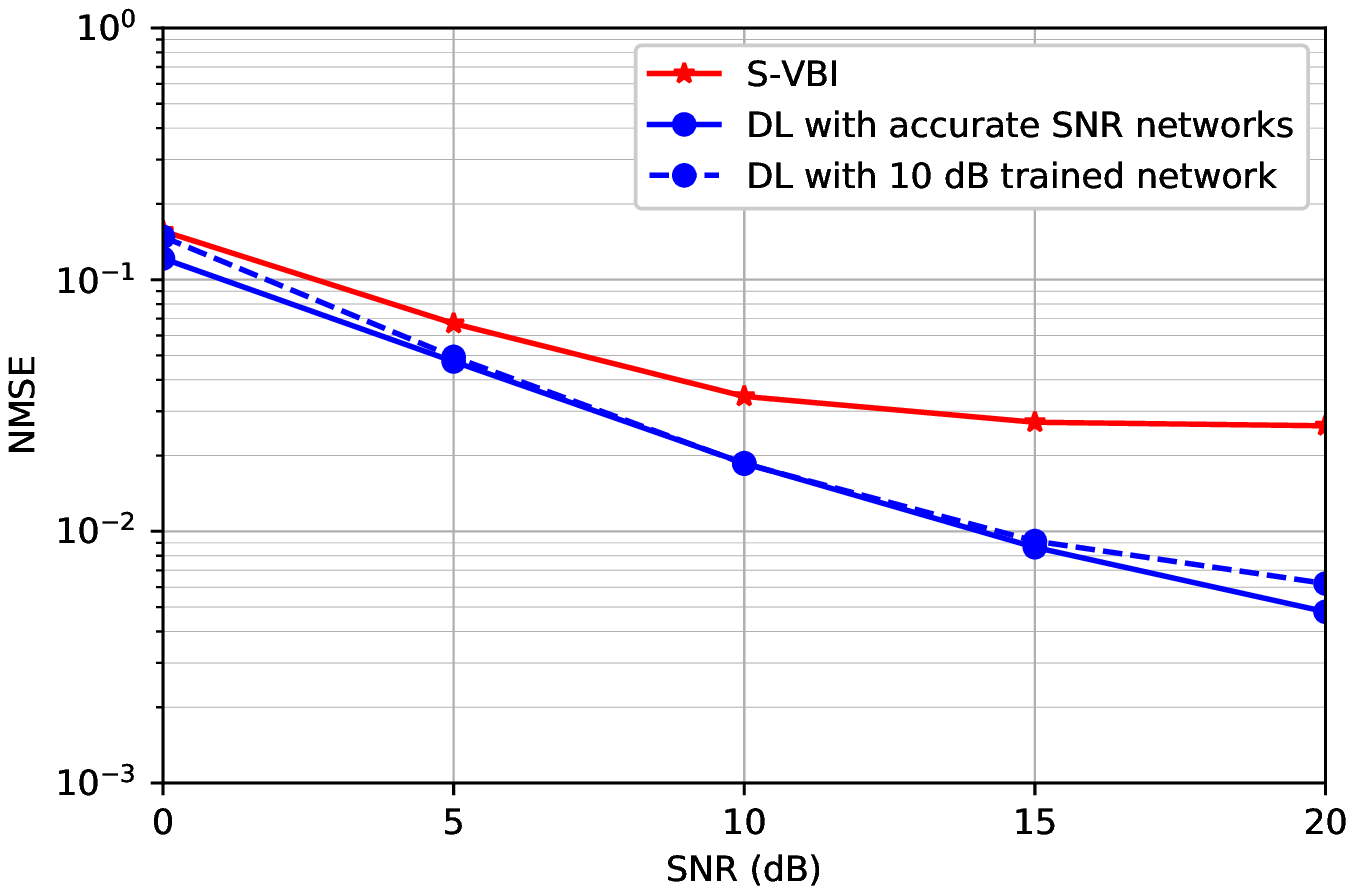}}
	\caption{The impact of SNR.}
	\label{impact_of_snr}
\end{figure}

The impact of SNR is illustrated in Fig. \ref{impact_of_snr}. It is straightforward that higher SNR results in better performance in general because of reduced noise effects. However, with 1/8 RF chains, the NMSE of S-VBI meets its floor when SNR increases above 15 dB, where the error is dominated by the limited ability of reversing the compression effects. In contrast, the NMSE of DL-based approach continues to decrease smoothly, which indicates its superiority. For the DL-based approach, the generalization to different testing SNRs is investigated as well. As can be indicated from the close curves with circle markers, testing with the network trained with only 10 dB data can achieve similar performance as testing with networks trained with accurate SNR data, especially when the SNR mismatch is not quite large. There are mainly two reasons. First, for a certain angular region, the learned measurement matrix is mainly determined by the distribution characteristics of the angular domain channels, therefore can be shared under different SNRs. Second, the DL-based channel estimator realizes functions of compression reversion and channel denoising. The former function is basically independent of noise, while the latter function can lead to slight performance degradation due to different denoising intensity under different SNRs. Therefore, we only need to train a single network with a moderate SNR in practice, then other SNRs can be properly handled.

\section{Conclusion}
In this paper, the channel estimation problem for HAD massive MIMO systems has been investigated and a DL-based framework has been proposed to solve this challenging problem. To exploit the spatial-clustered sparsity structure of angular domain channels, an angular space segmentation method has been developed to segment the entire angular space into multiple small angular regions, where a dedicated neural network containing region-specific measurement matrix and channel estimator is trained offline for each region. During online deployment, the best matching network is selected for the user based on its GPS information. Simulation results show that the proposed DL-based approach is superior to the state-of-the-art CS algorithms and the conventional measurement matrix in terms of both NMSE performance and computational complexity, providing a promising real-time solution for HAD massive MIMO systems.

%\newpage
\nocite{*}
\bibliographystyle{IEEE}
\begin{footnotesize}

\end{footnotesize}

\end{document}